\documentclass[onecolumn,smallextended]{svjour2}

\usepackage{graphicx}
\usepackage{dcolumn}
\usepackage{bm}
\usepackage{mdwlist}
\usepackage{setspace}
\usepackage{amssymb}
\usepackage{amsmath}
\usepackage{paralist}
\usepackage{enumitem}
\usepackage{fouridx}
\usepackage{overpic}
\usepackage{cases}
\usepackage{rotating}
\usepackage{color}
\usepackage{arydshln}
\usepackage{pst-node}
\usepackage{esint}
\usepackage[numbers,sort&compress]{natbib}

\DeclareMathOperator{\res}{Res}
\DeclareMathOperator{\Gam}{\Gamma}

\begin{document}

\title{Classical formulations of the electromagnetic self-force of extended charged bodies}

\author{P.~W.~Smorenburg \and L.~P.~J.~Kamp \and O.~J.~Luiten}
\institute{P.~W.~Smorenburg \and L.~P.~J.~Kamp \and O.~J.~Luiten \at Coherence and Quantum Technology (CQT), Eindhoven University of Technology, PO Box 513, 5600 MB Eindhoven, The Netherlands\\
Tel.: +31-40-2474048, Fax: +31-40-2438060, \email{o.j.luiten@tue.nl}}

\maketitle

\begin{abstract}
Several noncovariant formulations of the electromagnetic self-force of extended charged bodies, as have been developed in the context of classical models of charged particles, are compared. The mathematical equivalence of the various dissimilar self-force expressions is demonstrated explicitly by deriving these expressions directly from one another. The applicability of the self-force formulations and their significance in the wider context of classical charged particle models are discussed.
\end{abstract}

\section{Introduction \label{2sec1}}
The self-consistent description of charged particles is no doubt the longest-standing fundamental problem in classical electrodynamics. The conceptual difficulties, which invariably arise from the interaction of the particle with its self-generated electromagnetic fields, depend on the particle model that is postulated \cite{Jimenez}. In line with modern particle physics, often a structureless point charge model is adopted, leading to the well-known Lorentz-Abraham-Dirac (LAD) equation of motion \cite{Dirac,Landau} for elementary charged particles. As is also well known \cite{Panofsky}, however, the LAD equation is plagued by irreconcilable deficiencies manifested by runaway solutions, in which the particle momentum grows exponentially toward infinity, or by preacceleration solutions, where the particle starts to accelerate even before the onset of any external force. A popular remedy is to express in the LAD equation the particle acceleration perturbatively in terms of the external force \cite{Landau}, which renders the equation stable. However, arguably this procedure does nothing to improve the LAD equation intrinsically. It has been put forward that the stabilized equation may be considered as more fundamental than the LAD equation itself, but the argument usually requires abandoning the point particle limit proper by ascribing some structure to the particle \cite{Ford,Rohrlich2002,Aresdeparga}. Meanwhile, alternative classical electrodynamics of point particles continue to be proposed \cite{Hammond,Kholmetskii,Aresdeparga,Gill,Oliver,Villarroel,Bosanac}.\\

Another well-studied possibility in the development of a consistent charged particle electrodynamics is to dispense with the point charge model and associated problems altogether, and picture charged particles as extended charged bodies. In fact, this was historically the first charged particle model that was investigated \cite{Janssen,Lorentz,Abraham}. If the particle size is larger than a very small but finite critical length (which is comparable to the classical electron radius), the particle dynamics is free of the unphysical runaway solutions \cite{Wildermuth,Moniz,Medina}. This continues to motivate detailed calculations of the electromagnetic fields and forces inside accelerated extended charged bodies \cite{Roa-Neri2002,Roa-Neri1993,Medina,Yaghjian}. Moreover, extended charged particles keep open the possibility of electromagnetic interpretations of inertia \cite{Martins}, as originally proposed in pre-relativistic times \cite{Lorentz,Abraham}. Furthermore, the unphysical prediction of singular or preacceleration behavior of charged particles at a sudden onset of applied forces can be successfully removed by taking into account the finite propagation velocity of signals traversing the extended particle \cite{Yaghjian}. The latter effect is the classical analogue of the concept of "self-dressing" familiar from quantum electrodynamics \cite{Compagno}, and plays a role in the measurability of the electromagnetic field \cite{Hnizdo}. Finally, and perhaps most significantly, extended charged particle models provide an important way to access point particle models by taking the appropriate limit corresponding to vanishing particle radius. Consequently, the physical and mathematical consistency of this limiting procedure is itself subject of active research \cite{Gralla,Aguirregabiria,Noja}.\\

\begin{figure*}[t]
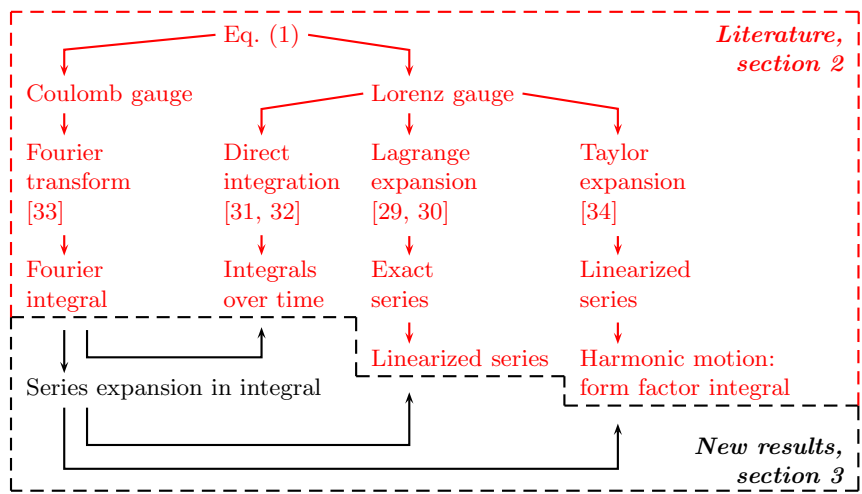

\vspace{3mm}
\setlength{\tabcolsep}{1mm}
\addtolength{\abovecaptionskip}{-3mm}
\centering
\small{
\begin{tabular}{lllllll}
\multicolumn{1}{l}{\Rnode[vref=1mm]{x}{}}&&\Rnode{a}{\color{red}{Eq. (\ref{2.1})}}&&&&\multicolumn{1}{r}{\color{red}{\textbf{\emph{Literature,}}}\Rnode[vref=1mm]{w}{}}\\
&&&&&&\multicolumn{1}{r}{\color{red}{\textbf{\emph{section \ref{2sec2}}}}}\\
\Rnode[href=-1]{b}{\color{red}{Coulomb gauge}}&&&&\Rnode[href=-1]{f}{\color{red}{Lorenz gauge}}&&\\
&&&&&&\\
\Rnode[href=-1]{c}{\color{red}{Fourier}}&&\Rnode[href=-1]{g}{\color{red}{Direct}}&&\Rnode[href=-1]{i}{\color{red}{Lagrange}}&&\Rnode[href=-1]{l}{\color{red}{Taylor}}\\
\color{red}{transform}&&\color{red}{integration}&&\color{red}{expansion}&&\color{red}{expansion}\\
\color{red}{\cite{Bohm}}&&\color{red}{\cite{Sommerfeld1,Sommerfeld2}}&&\color{red}{\cite{Herglotz,Schott}}&&\color{red}{\cite{Jackson}}\\
&&&&&&\\
\Rnode[href=-1]{d}{\color{red}{Fourier}}&&\Rnode[href=-1]{h}{\color{red}{Integrals}}&&\Rnode[href=-1]{j}{\color{red}{Exact}}&&\Rnode[href=-1]{m}{\color{red}{Linearized}}\\
\color{red}{integral}&&\color{red}{over time}&&\color{red}{series}&&\color{red}{series}\\
\Rnode[vref=2.5mm]{v}{}&&&\Rnode[vref=2.5mm]{u}{}&&\\
&&&&\Rnode[href=-1]{k}{\color{red}{Linearized series}}&&\Rnode[href=-1]{n}{\color{red}{Harmonic motion:}}\\
\multicolumn{3}{l}{\Rnode[href=-1]{e}{Series expansion in integral}}&\Rnode[vref=2.5mm]{t}{}&&\Rnode[vref=2.5mm]{s}{}&\Rnode[href=-1]{na}{\color{red}{form factor integral}}\\
&&&&&\multicolumn{1}{r}{\Rnode[vref=2.5mm]{r}{}}&\multicolumn{1}{r}{\Rnode[vref=2.5mm]{o}{}}\\
&&&&&&\multicolumn{1}{r}{\hspace{1.5cm}\textbf{\emph{New results,}}}\\
\Rnode[vref=-1mm]{q}{}&&&&&&\multicolumn{1}{r}{\textbf{\emph{section \ref{2sec3}}}\Rnode[vref=-1mm]{p}{}}\\
\ncdiag[nodesep=3pt,angleB=90,armA=0mm,angleA=180,offsetB=5mm,linecolor=red]{->}{a}{b}
\ncline[nodesep=3pt,offset=5mm,linecolor=red]{->}{b}{c}
\ncline[nodesepA=10mm,nodesepB=2pt,offset=5mm,linecolor=red]{->}{c}{d}
\ncline[nodesepA=7mm,nodesepB=2pt,nodesepB=2pt,offset=5mm]{->}{d}{e}
\ncdiag[nodesep=3pt,angleB=90,armA=0mm,offsetB=5mm,linecolor=red]{->}{a}{f}
\ncdiag[nodesep=3pt,angleB=90,armA=0mm,offsetB=5mm,angleA=180,linecolor=red]{->}{f}{g}
\ncline[nodesepA=10mm,nodesepB=2pt,offset=5mm,linecolor=red]{->}{g}{h}
\ncline[nodesep=3pt,offset=5mm,linecolor=red]{->}{f}{i}
\ncline[nodesepA=9.5mm,nodesepB=2pt,offset=5mm,linecolor=red]{->}{i}{j}
\ncline[nodesepA=6mm,nodesepB=2pt,offset=5mm,linecolor=red]{->}{j}{k}
\ncdiag[nodesep=3pt,armA=0mm,angleB=90,offsetB=5mm,linecolor=red]{->}{f}{l}
\ncline[nodesepA=9.5mm,nodesepB=2pt,offset=5mm,linecolor=red]{->}{l}{m}
\ncline[nodesepA=6mm,nodesepB=2pt,offset=5mm,linecolor=red]{->}{m}{n}
\ncangle[nodesepA=3pt,nodesepB=9pt,angleA=-90,angleB=-90,offsetA=5mm,offsetB=-5mm,armB=6mm]{->}{e}{na}
\ncangle[nodesepA=3pt,nodesepB=10pt,angleA=-90,angleB=-90,offsetA=8mm,offsetB=-5mm,armB=7mm]{->}{e}{k}
\ncangle[nodesepA=7mm,nodesepB=6mm,angleA=-90,angleB=-90,offsetA=8mm,offsetB=-5mm,armB=4mm]{->}{d}{h}
\ncline[nodesep=0pt,linestyle=dashed,offset=2mm]{-}{q}{v}
\ncline[nodesepB=0pt,nodesepA=-2mm,linestyle=dashed]{-}{v}{u}
\ncline[nodesep=0pt,linestyle=dashed]{-}{u}{t}
\ncline[nodesep=0pt,linestyle=dashed]{-}{t}{s}
\ncline[nodesep=0pt,linestyle=dashed]{-}{s}{r}
\ncline[nodesep=0pt,linestyle=dashed,nodesepB=-2mm]{-}{r}{o}
\ncline[nodesep=0pt,linestyle=dashed,offset=2mm]{-}{o}{p}
\ncline[nodesepA=-2mm,nodesepB=-2mm,linestyle=dashed]{-}{p}{q}
\ncline[linestyle=dashed,offset=-2mm,linecolor=red,nodesepB=-4mm,nodesepA=0pt]{-}{o}{w}
\ncline[nodesepA=-2mm,nodesepB=-2mm,linestyle=dashed,linecolor=red,offset=-3mm]{-}{w}{x}
\ncline[nodesepB=0pt,linestyle=dashed,linecolor=red,offset=-2mm,nodesepA=-4mm]{-}{x}{v}
\end{tabular}}
\caption{The content of this paper in relation to existing results in literature}
\label{2fig1}
\end{figure*}

For the above reasons, a thorough understanding of the electrodynamics of extended charged bodies is a prerequisite for the development of classical charge particle models of both the point charge and the extended charge variety. Spread out over a century, however, many formulations of these dynamics have appeared in often dissimilar forms, making a broad comparison of results difficult. In this paper, we aim to contribute to a more coherent picture of extended particle models by comparing several published expressions \cite{Herglotz,Schott,Sommerfeld1,Sommerfeld2,Bohm,Jackson} for the self-force of a rigid charged body. We demonstrate the equivalence of these dissimilar expressions by deriving them directly from one another. The self-force $\bm{F}$, which is the resultant Lorentz force that is experienced by the charged body and caused by the self-produced electromagnetic fields, reads in noncovariant form
\begin{align}
\bm{F}=\int\left(\rho\bm{E}+\bm{J}\times\bm{B}\right)d^3\bm{x}=\!\int\!\left[-\rho\left(\nabla\phi+\frac{\partial\bm{A}}{\partial t}\right)+\bm{J}\times\left(\nabla\times\bm{A}\right)\right]\!d^3\bm{x},\label{2.1}
\end{align}
where $\rho$ is the charge density, $\bm{J}$ is the current density, $\bm{E}$ is the electric field, $\bm{B}$ is the magnetic field, $\phi$ and $\bm{A}$ are the electromagnetic potentials, and the integration is over the extent of the charged body. The self-force calculations that will be considered here take Eq. (\ref{2.1}) as a starting point, and have the form of series expansions \cite{Herglotz,Schott}, definite integrals over retarded time \cite{Sommerfeld1,Sommerfeld2}, and Fourier integrals \cite{Bohm,Jackson}. Figure \ref{2fig1} shows schematically the content of this paper in relation to these publications. It should be stressed that the figure represents only a very small fraction of the available literature on the subject; correspondingly this paper is not meant as a comprehensive review. Rather, the new connections that will be established here complete the literature shown in Fig. \ref{2fig1}, and present the results on a common basis. Section \ref{2sec2} gives an overview of the considered existing self-force expressions. In section \ref{2sec3} it is shown how the various expressions follow directly from each other. Section \ref{2sec4} discusses the applicability of the self-force formulations, and puts them in the wider context of classical charge particle models. Like the primary equation (\ref{2.1}), our formulation will be noncovariant throughout.
\renewcommand{\arraystretch}{1.2}
\renewcommand{\arraystretch}{1}

\section{Existing self-force derivations \label{2sec2}}
The derivation of the self-force of a rigid charged body requires the evaluation of Eq. (\ref{2.1}) by some method. This involves a calculation of the electromagnetic potentials appearing in Eq. (\ref{2.1}), which necessitates a choice of gauge. The potentials $\phi^{(L)}$ and $\bm{A}^{(L)}$ in the Lorenz gauge and the vector potential $\bm{A}^{(C)}$ in the Coulomb gauge satisfy the wave equations
\begin{align}
\left(\frac{1}{c^2}\frac{\partial^2}{\partial t^2}-\nabla^2\right)\Psi(\bm{x},t)=\mu_0c\Pi(\bm{x},t).\label{2.2}
\end{align}
Here, $\Psi\equiv\left\{\phi^{(L)},c\bm{A}^{(L)},c\bm{A}^{(C)}\right\}$ and $\Pi\equiv\left\{c\rho,\bm{J},\bm{J}_T\right\}$, with $\bm{J}_T$ the divergenceless part of the current density \cite{Jackson}. The scalar potential in the Coulomb gauge is not relevant to the problem, as will be discussed below. In terms of the causal Green's function for the wave equation, which equals \cite{Jackson}
\begin{align}
G(\bm{x},\bm{x}',t,t')=\frac{\delta\left(t-t'-\left|\bm{x}-\bm{x}'\right|/c\right)}{\left|\bm{x}-\bm{x}'\right|},\label{2.4}
\end{align}
with $\delta$ the Dirac delta function, the relevant particular solution of Eq. (\ref{2.2}) is given by
\begin{align}
\Psi(\bm{x},t)=\frac{\mu_0c}{4\pi}\iint G(\bm{x},\bm{x}',t,t')\Pi(\bm{x}',t')d^3\bm{x}'dt'.\label{2.5}
\end{align}
For any given charge distribution and given history of the motion of the charged body, $\Pi(\bm{x}',t')$ is known, so that in principle the potentials can be evaluated with Eq. (\ref{2.5}), after which the self-force can be determined via Eq. (\ref{2.1}). The calculations available in literature where this program is followed differ in the order in which the integrations in Eq. (\ref{2.5}) are carried out. In view of the delta function in Eq. (\ref{2.4}), it is tempting to start with the integration with respect to $t'$. This immediately yields the well-known retarded integral expressions \cite{Jackson} for the potentials, which indeed are the starting point for the self-force calculations presented in sections \ref{2sec2a} and \ref{2sec2b} below. However, integrating first with respect to $t'$ in Eq. (\ref{2.5}) is not the only possibility. For certain charge distributions, it is advantageous to start with the integration over $\bm{x}'$, as will be described in section \ref{2sec2c}. Still another possibility is to Fourier transform Eq. (\ref{2.5}), that is, to integrate with respect to the coordinates $\bm{x}$; this is shown in section \ref{2sec2d}.
\subsection{Taylor expansion \label{2sec2a}}
Adopting the Lorenz gauge, integration of Eq. (\ref{2.5}) with respect to $t'$ yields the retarded integral expressions
\begin{align}
\Psi(\bm{x},t)=\frac{\mu_0c}{4\pi}\int\frac{\Pi(\bm{x}',t_{ret})}{\left|\bm{x}-\bm{x}'\right|}d^3\bm{x}',\label{2.10}
\end{align}
where now $\Psi\equiv\left\{\phi,c\bm{A}\right\}$ and $\Pi\equiv\left\{c\rho,\bm{J}\right\}$. In Eq. (\ref{2.10}), the integration is complicated by the fact that $\Pi$ must be evaluated at the retarded time $t_{ret}\equiv t-\left|\bm{x}-\bm{x}'\right|/c$, which is different for each volume element $d^3\bm{x}'$. Jackson \cite{Jackson} approaches this problem by expanding $\Pi$ in a Taylor series around the current time $t$,
\begin{align}
\Pi(\bm{x}',t_{ret})=\sum_{n=0}^\infty\frac{1}{n!}\left(-\frac{\left|\bm{x}-\bm{x}'\right|}{c}\right)^n\frac{\partial^n \Pi(\bm{x}',t)}{\partial t^n}.\label{2.11}
\end{align}
Substitution of Eq. (\ref{2.11}) in Eq. (\ref{2.10}) expresses the potential in terms of quantities evaluated at the current time only. Using the result in Eq. (\ref{2.1}) gives, after some manipulations \cite{Jackson}, the electric part of the self-force
\begin{align}
\bm{F}=-\int\hspace{-1mm} &d^3\bm{r}\,\rho(\bm{r},t)\left[\frac{\mu_0}{4\pi}\sum_{n=0}^\infty\frac{(-1)^n}{n!c^n}\right.\label{2.11a}\\
&\left.\int\hspace{-1mm}R^{n-1}\frac{\partial^{n+1}}{\partial t^{n+1}}\left(\frac{n+1}{n+2}\,\bm{J}(\bm{r}',t)-\frac{n-1}{n+2}\frac{\left[\bm{J}(\bm{r}',t)\cdot\bm{R}\right]\bm{R}}{R^2}\right)d^3\bm{r}'\vphantom{\sum_{n=0}^\infty}\right].\nonumber
\end{align}
Here, the integration variables have been changed to $\bm{r}=\bm{x}-\bm{\xi}(t),\bm{r}'=\bm{x}'-\bm{\xi}(t)$ where $\bm{\xi}(t)$ is the trajectory of the center of the charged body, and $\bm{R}=\bm{r}-\bm{r}'$. For a spherically symmetric rigid charge distribution, Eq. (\ref{2.11a}) simplifies to
\begin{align}
\bm{F}=-\frac{\mu_0}{6\pi}\sum_{n=0}^\infty\frac{(-1)^n}{n!c^n}\frac{d^{n+2}\bm{\xi}}{dt^{n+2}}\iint\rho(\bm{r})\rho(\bm{r}')\left|\bm{r}-\bm{r}'\right|^{n-1}d^3\bm{r}d^3\bm{r}'.\label{2.12}
\end{align}
In Eq. (\ref{2.12}), the magnetic part corresponding to the last term in square brackets in Eq. (\ref{2.1}) has been neglected, so that Eq. (\ref{2.12}) is the self-force linearized in $\bm{\xi}$ and its time derivatives. In case of harmonic motion $\bm{\xi}=\bm{\xi}_0\exp(-i\omega t)\equiv\widetilde{\bm{\xi}}$, the series in Eq. (\ref{2.12}) can be readily summed, and is proportional to $\exp(i\omega\left|\bm{r}-\bm{r}'\right|/c)$. Furthermore, writing in Eq. (\ref{2.12}) the charge distributions $\rho$ in terms of their spatial Fourier transforms, and integrating the resulting expression, it is found that \cite{Jackson}
\begin{align}
\bm{F}=\frac{8\pi\omega^2}{3\epsilon_0c^2}\,\widetilde{\bm{\xi}}\,\lim_{\lambda\downarrow 0}\int_0^\infty\hspace{-2mm}\frac{k^2\left|\rho_k\right|^2}{k^2-(\omega/c+i\lambda)^2}dk.\label{2.13}
\end{align}
Here, the symmetrical convention for Fourier transformed quantities $\bm{Y}_k\equiv(2\pi)^{-3/2}\int\bm{Y}(\bm{r})\exp(-i\bm{k}\cdot\bm{r})d^3\bm{r}$ is adopted. The quantity $\rho_k$ is often called the form factor of the charge distribution.
\subsection{Lagrange expansion \label{2sec2b}}
As mentioned above, Eq. (\ref{2.12}) is a linearized approximation to the exact self-force due to the neglect of the magnetic term in Eq. (\ref{2.1}). However, the derivation in section \ref{2sec2a} is inexact for another reason. Namely, by making use of a predefined rigid charge distribution $\rho$ throughout the derivation (or more precisely, using the distribution in the proper frame), it is implied that the potentials are generated by a total charge $\int\rho(\bm{x}',t_{ret})d^3\bm{x}'$. The latter is in general not equal to the true charge of the body $\int\rho(\bm{x}',t)d^3\bm{x}'\equiv q$, but rather depends on the body's state of motion. To correct for this inconsistency, either the quantity $\Pi$ should be defined in a relativistically covariant way, or else the integral (\ref{2.10}) should be modified to leave the total charge invariant. The latter, however, is precisely how the Li\'enard-Wiechert potentials for a moving point charge were devised, as is explained clearly in Ref. \cite[sec. 19.1]{Panofsky}. Accordingly, the charged body may be regarded as a collection of infinitesimal particles moving with the trajectory $\bm{\xi}(t)+\bm{r}'$ and having a fixed charge $\rho(\bm{r}')d^3\bm{r}'$ with $\rho$ the proper frame distribution. The corresponding potentials are thus given by
\begin{align}
\Psi(\bm{x},t)=\frac{\mu_0c}{4\pi}\int\left.\frac{\left\{c,\bm{v}\right\}}{R-\bm{R}\cdot\bm{v}/c}\right|_{t=t_{ret}}\hspace{-2mm}\rho(\bm{r'})d^3\bm{r}',\label{2.30}
\end{align}
where $\bm{v}(t)=d\bm{\xi}/dt$ is the velocity of the charged body, and $\bm{R}(t)\equiv\bm{x}-\bm{\xi}(t)-\bm{r}'$. An important difference between Eq. (\ref{2.10}) and the Li\'enard-Wiechert formulation Eq. (\ref{2.30}), apart from the different denominator, is that in the former the retarded time was known explicitly in terms of the coordinates $\bm{x}$ and $\bm{x}'$, while in the latter it is only defined implicitly by the retardation condition $t_{ret}=t-R(t_{ret})/c$. This complicates the derivation of the self-force significantly. Herglotz \cite{Herglotz} and Schott \cite{Schott} proceeded by expanding retarded quantities $Y$ in series using Lagrange's reversion theorem \cite{Whittaker},
\begin{align}
Y(t_{ret})=Y(t)+\sum_{n=1}^\infty\frac{(-1)^n}{n!\,c^n}\frac{d^{n-1}}{dt^{n-1}}\left[R(t)^n\frac{dY(t)}{dt}\right].\label{2.31}
\end{align}
Note that differentiation of the quantity $R^n$ in Eq. (\ref{2.31}) produces factors of the velocity $\bm{v}$ and derivatives thereof, so that the Taylor series Eq. (\ref{2.11}) is in fact a linearization of Eq. (\ref{2.31}) in which all terms nonlinear in $\bm{v}$ and its derivatives have been neglected. Likewise, the potentials Eq. (\ref{2.10}) are linearizations of the Li\'enard-Wiechert potentials Eq. (\ref{2.30}). Now let $b$ be the characteristic size of the charged body. On working out the first few terms of Eq. (\ref{2.31}), and noting that $R\sim b$ for relevant field points $\bm{x}$, it becomes apparent that these linearizations are good approximations provided that
\begin{align}
\left|\frac{b^n}{c^n}\frac{d^n}{dt^n}\bm{v}\right|\ll\left|\bm{v}\right|\label{2.31a}
\end{align}
for $n\geq1$. Roughly speaking, this means that the motion of the body should not change significantly on the time scale necessary for light to travel across the body, which is the time scale at which self-forces are communicated. This condition is known as quasi-stationary motion \cite{Erber}. It indicates the range of validity of the form factor integral Eq. (\ref{2.13}), in addition to the condition $\left|\bm{v}\right|\ll c$ associated with the neglect of magnetic forces.\\

Substitution of Eq. (\ref{2.31}) in Eq. (\ref{2.30}) expresses the potentials in terms of quantities evaluated at the current time only. Using the result in Eq. (\ref{2.1}), and performing all integrations, gives a series expansion for the self-force. This series has been evaluated explicitly up to cubic terms in the velocity for a homogeneously charged sphere of radius $b$ by Herglotz \cite{Herglotz}. The linear terms are
\begin{align}
\bm{F}=-\frac{6\mu_0q^2}{\pi b}\sum_{n=0}^\infty\frac{(n+1)(n+4)(-2b/c)^n}{(n+5)!}\frac{d^{n+2}\bm{\xi}}{dt^{n+2}},\label{2.32}
\end{align}
and dominate the nonlinear terms in case of quasi-stationary motion Eq. (\ref{2.31a}). For a homogeneously charged sphere in rectilinear motion $\bm{\xi}(t)=\xi(t)\bm{e}_z$, Schott \cite{Schott} derived the following closed-form expression including terms up to arbitrary order:
\begin{align}
\bm{F}=-\frac{36q^2}{\pi\epsilon_0b}\,\bm{e}_z\sum_{n=0}^\infty\sum_{m=0}^\infty&\frac{(m+1)(n+1)(n+4)(-2b)^n}{(2m+1)(2m+3)!(n+5)!}\label{2.33}\\
&\times\left.\frac{\partial^{2m+n+2}}{\partial u^{2m+n+2}}\left[\xi(t+u/c)-\xi(t)\right]^{2m+1}\right|_{u=0},\nonumber
\end{align}
which reduces to Eq. (\ref{2.32}) when truncated at $m=0$.
\subsection{Direct evaluation \label{2sec2c}}
Sommerfeld \cite{Sommerfeld1,Sommerfeld2} has evaluated the self-generated potentials of a charged body by integrating Eq. (\ref{2.5}) with respect to $\bm{x}'$. For a homogeneously charged sphere, this leaves a one-dimensional integral over $t'$ \cite{Sommerfeld2}:
\begin{align}
\Psi(\bm{x},t)=-\frac{3q}{16\pi^2b^3}\int_{-\infty}^t\frac{\left\{c,\bm{v}(t')\right\}}{R_c(t')}\chi\,dt',\label{2.50}
\end{align}
where $R_c(t')=\left|\bm{x}-\bm{\xi}(t')\right|$ is the distance to the center of the sphere, and
\begin{align}
\chi=
\begin{cases}
4c(t-t')R_c&\hspace{5mm}c(t-t')<b-R_c\\
b^2-\left[c(t-t')-R_c\right]^2&\hspace{5mm}b-R_c<c(t-t')<b+R_c\\
0&\hspace{5mm}c(t-t')>b+R_c
\end{cases}.
\label{2.51}
\end{align}
In virtue of the delta function in Eq. (\ref{2.5}), times between $t'$ and $t'+dt'$ in Eq. (\ref{2.50}) correspond to the contribution to the potentials $\Psi(\bm{x},t)$ generated by the charge located within a shell with radius $c(t-t')$ and thickness $cdt'$ centered around the field point $\bm{x}$. Depending on $R_c$ and $t'$, this shell may fall completely within the charged sphere, or only partially, or not at all, each case leading to a different factor $\chi$ as given by Eq. (\ref{2.51}). Using the potentials Eq. (\ref{2.50}) in Eq. (\ref{2.1}) results in the self-force \cite{Sommerfeld1}
\begin{align}
\bm{F}=-\frac{3q^2}{32\pi\epsilon_0b^4c}\left(\int_0^{\tau^+}\hspace{-2mm}G^+(\tau)\,d\tau-\int_0^{\tau^-}\hspace{-2mm}G^-(\tau)\,d\tau\right),\label{2.52}
\end{align}
in which the integrations are over the time difference $\tau\equiv t-t'$, and
\begin{align*}
G^\pm(\tau)&=\left[c^2-\bm{v}(t)\cdot\bm{v}(t-\tau)\right]\frac{\bm{s}}{s}\frac{\partial}{\partial s}\frac{g(c\tau\pm s)}{s}+\frac{\partial}{\partial t}\frac{\bm{v}(t-\tau)g(c\tau\pm s)}{s}\,;\\
g(y)&=\frac{y^5}{20b^5}-y^3+2by^2-\frac{8b^3}{5}.
\end{align*}
Here, $\bm{s}=\bm{\xi}(t)-\bm{\xi}(t-\tau)$ is the displacement of the charged sphere during the time interval $\tau$. The upper integration limits in Eq. (\ref{2.52}) are the roots of the equations $c\tau^\pm\pm s(\tau^\pm)=2b$. These limits demarcate different stages in the communication of electromagnetic signals between the parts of the charged sphere that lead to the self-force at the current time $t$. For subluminal motion, the trailing end of the sphere receives electromagnetic signals at time $t$ that were emitted by the other parts of the sphere at times between $t-\tau^+$ and $t$. The signals received by the leading end at time $t$ were emitted by the other parts during the slightly longer interval between $t-\tau^-$ and $t$. Signals emitted at still earlier at times before $t-\tau^-$ do not arrive at any other part of the sphere at time $t$, so that the domain $\tau>\tau^-$ does not contribute to the self-force Eq. (\ref{2.52}) at all.
\subsection{Fourier transform \label{2sec2d}}
Bohm and Weinstein \cite{Bohm} have adopted the Coulomb gauge to evaluate Eq. (\ref{2.1}). The benefit of this gauge choice for the calculation of the self-force of a rigid charged body is that the scalar potential $\phi$ equals the electrostatic potential corresponding to the instantaneous distribution of charge. Since for any pair of charge elements $de_1$ and $de_2$ the instantaneous electrostatic force on $de_1$ due to $de_2$ is the negative of the electrostatic force on $de_2$ due to $de_1$, the contribution of $\phi$ to the self-force $\bm{F}$ integrates to zero identically. Therefore only the vector potential has to be taken into account, which is given by Eq. (\ref{2.5}) as before. It can be shown \cite{Bohm} that a Fourier transformation of this equation from the spatial domain $\bm{x}$ to the wave vector domain $\bm{k}$ yields the potential
\begin{align}
\bm{A}_k(\bm{k},t)=\frac{\mu_0c}{k}\int_{-\infty}^t\bm{J}_{T,k}(\bm{k},t)\sin\left[ck(t-t')\right]dt'.\label{2.100}
\end{align}
Notice that the integration in Eq. (\ref{2.100}) extends to the upper boundary $t$, so that the potential at time $t$ depends only on currents at past times $t'<t$, that is, Eq. (\ref{2.100}) is causal as it should be. Using in the self-force Eq. (\ref{2.1}) the inverse Fourier transform $\bm{A}\equiv(2\pi)^{-3/2}\int\bm{A}_k\exp(i\bm{k}\cdot\bm{x})d^3\bm{k}$, and substituting Eq. (\ref{2.100}), gives \cite{Bohm}
\begin{align}
\bm{F}=-\frac{1}{\epsilon_0}\int_{-\infty}^t\hspace{-2mm}dt'&\int d^3\bm{k}\left|\rho_k\right|^2\exp\left(i\bm{k}\cdot\bm{s}\right)\label{2.101}\\
&\times\left(\frac{\bm{k}\times[\bm{v}(t')\times\bm{k}]}{k^2}\cos ck\tau-\frac{\bm{v}(t)\times[\bm{k}\times\bm{v}(t')]}{ck}i\sin ck\tau\right).\nonumber
\end{align}
Note that the second term in large braces is proportional to and perpendicular to the current velocity $\bm{v}(t)$ of the charged body, and therefore represents the magnetic component of the self-force. The first term gives the electric component. For a spherically symmetric charge distribution, $\rho_k(\bm{k})$ is a function of the magnitude of $\bm{k}$ but not of its direction. In this case, Eq. (\ref{2.101}) can be straightforwardly integrated over angles in $\bm{k}$-space. This reduces Eq. (\ref{2.101}) to
\begin{align}
&\hspace{-1mm}\bm{F}=-\frac{4\pi}{\epsilon_0}\int_{-\infty}^t\hspace{-2mm}dt'\int_0^\infty \hspace{-2mm}dk\,k^2\left|\rho_k\right|^2\left[\left(\bm{v}(t')-\frac{[\bm{v}(t')\cdot\bm{s}]\bm{s}}{s^2}\right)j_0\left(ks\right)\cos ck\tau\right.\label{2.102}\\
&\hspace{-1mm}\left.-\left(\bm{v}(t')-\frac{3[\bm{v}(t')\cdot\bm{s}]\bm{s}}{s^2}\right)\frac{j_1\left(ks\right)}{ks}\cos ck\tau+\frac{\bm{v}(t)\times[\bm{s}\times\bm{v}(t')]}{cs}j_1\left(ks\right)\sin ck\tau\right]\!,\nonumber
\end{align}
where $j_n$ denotes the spherical Bessel function of the first kind and order $n$ \cite{Abramowitz}. In Eq. (\ref{2.102}), the first two terms in large square brackets represent the electric component of the force and are given in Ref. \cite{Bohm}; the last term gives the magnetic component. The integral over $k$ containing Bessel function kernels has the typical form of an inverse Fourier transform in spherical coordinates \cite{Stratton}. In section \ref{2sec3}, we will derive the other self-force representations given in sections \ref{2sec2a} to \ref{2sec2c} from this Fourier integral.\\

\section{Equivalence of the self-force expressions \label{2sec3}}
\subsection{Fourier integral and integral over time \label{2sec3a}}
Sommerfeld derived for the piecewise function $\chi$ given by Eq. (\ref{2.51}) the integral representation \cite{Sommerfeld2}
\begin{align}
\chi=\frac{8b^2}{\pi}\int_0^\infty\hspace{-1mm}\frac{j_1(kb)\sin(kR_c)\sin[ck(t-t')]}{k}\,dk.\label{2.110}
\end{align}
Substituting this representation in Eq. (\ref{2.50}), and using the result in Eq. (\ref{2.1}), gives the self-force \cite{Sommerfeld2}
\begin{align}
\bm{F}&=-\frac{9q^2}{2\pi^2\epsilon_0b^2c}\int_0^\infty\hspace{-2mm}d\tau\int_0^\infty\hspace{-2mm} dk\left[j_1(kb)\right]^2\label{2.111}\\
&\times\left(\frac{1}{k}\frac{\partial}{\partial t}\left[\bm{v}(t-\tau)j_0(ks)\sin ck\tau\right]-\left[c^2-\bm{v}(t)\cdot\bm{v}(t-\tau)\right]\frac{\bm{s}}{s}j_1(ks)\sin ck\tau\vphantom{\frac{\partial}{\partial t}}\right).\nonumber
\end{align}
Performing the integration over $k$ indeed yields the force Eq. (\ref{2.52}). We now show that Eq. (\ref{2.111}) is equivalent to Eq. (\ref{2.102}) that was derived by Fourier analysis of the potentials in the Coulomb gauge. Note that the integrands of both equations already have a similar structure due to the form of the integral representation Eq. (\ref{2.110}). Performing the differentiation $\partial/\partial t$ in Eq. (\ref{2.111}) using the property
\begin{align}
\frac{\partial j_0(ks)}{\partial t}=-kj_1(ks)\frac{\partial s}{\partial t}=-kj_1(ks)\frac{\left[\bm{v}(t)-\bm{v}(t-\tau)\right]\cdot\bm{s}}{s}\nonumber
\end{align}
gives, after changing the integration variable back to $t'$ and rearranging,
\begin{align}
\bm{F}&=-\frac{9q^2}{2\pi^2\epsilon_0b^2c}\int_{-\infty}^t\hspace{-2mm}dt'\int_0^\infty\hspace{-2mm} dk\,\left[j_1(kb)\right]^2\sin ck(t-t')\left[\frac{c^2\bm{s}}{s}j_1(ks)-\right.\label{2.113}\\
&\left.-\frac{1}{k}\frac{d\bm{v}(t')}{dt'}j_0(ks)+\left(\frac{\left\{\left[\bm{v}(t)-\bm{v}(t')\right]\cdot\bm{s}\right\}\bm{v}(t')}{s}-\frac{\left[\bm{v}(t)\cdot\bm{v}(t')\right]\bm{s}}{s}\right)j_1(ks)\right].\nonumber
\end{align}
Next we integrate by parts the first two terms in the large square brackets with respect to $t'$, choosing for the differentiated factors respectively $f_1(t')=\bm{s}j_1(ks)/s$ and $f_2(t')=j_0(ks)\sin ck\tau$. To carry out this integration unambiguously, it is necessary to replace the lower integration limit $t'=-\infty$ by $t'=-a$, and take the limit $a\rightarrow\infty$ afterwards. With the help of the relations
\begin{align*}
\frac{\partial f_1}{\partial t'}&=\left(\frac{3j_1(ks)}{ks}-j_0(ks)\right)\frac{k\left[\bm{v}(t')\cdot\bm{s}\right]\bm{s}}{s^2}-\frac{\bm{v}(t')}{s}j_1(ks);\\
\frac{\partial f_2}{\partial t'}&=\frac{k\bm{v}(t')\cdot\bm{s}}{s}j_1(ks)\sin ck\tau-ckj_0(ks)\cos ck\tau,
\end{align*}
the resulting self-force is
\begin{align}
\bm{F}&=\frac{9q^2}{2\pi^2\epsilon_0b^2}\lim_{a\rightarrow\infty}\int_0^\infty\hspace{-2mm}\left[j_1(kb)\right]^2\left(B+I\,\right)dk,\label{2.116}
\end{align}
where
\begin{align*}
&B=\left[\frac{\bm{v}(t')}{ck}j_0(ks)\sin ck\tau-\frac{\bm{s}}{ks}j_1(ks)\cos ck\tau\right]_{t'=-a}^t;\\
&I=\int_{-a}^t\left[\left(\bm{v}(t')-\frac{[\bm{v}(t')\cdot\bm{s}]\bm{s}}{s^2}\right)j_0\left(ks\right)\cos ck\tau\right.\\
&\!\!\left.-\left(\bm{v}(t')-\!\frac{3[\bm{v}(t')\cdot\bm{s}]\bm{s}}{s^2}\right)\!\frac{j_1\left(ks\right)}{ks}\cos ck\tau+\!\frac{\bm{v}(t)\times[\bm{s}\times\bm{v}(t')]}{cs}j_1\left(ks\right)\sin ck\tau\right]\!dt'\vphantom{\left[\frac{\bm{v}(t')}{ck}\right]_{t'=-\infty}^t}\!.
\end{align*}
Taking in Eq. (\ref{2.116}) the limit $a\rightarrow\infty$ of $I$ presents no difficulties, and yields precisely Eq. (\ref{2.102}), specialized to a homogeneous sphere that has the form factor
\begin{align}
\rho_k=\frac{3q}{(2\pi)^{3/2}}\frac{j_1(kb)}{kb}.\label{2.117}
\end{align}
Therefore Eq. (\ref{2.116}) is equivalent to Eq. (\ref{2.102}), provided that the boundary term $B$ vanishes. This can be shown to be the case as follows. $B$ evaluated at $t'=t$ vanishes since $\sin ck\tau=0$ and $s(t')=0$ at $t'=t$. In the limit ${t'\rightarrow-\infty}$, the first term of $B$ is zero trivially when $\bm{v}(-\infty)=\bm{0}$. When $\bm{v}(-\infty)\neq\bm{0}$, it must be that $s(t')\rightarrow\infty$ and hence $j_0(ks)\rightarrow0$ as $t'\rightarrow-\infty$, so that the first term does not contribute in this case either. The second term of $B$, on the other hand, vanishes at $t\rightarrow-\infty$ only when $\bm{v}(-\infty)\neq\bm{0}$. Namely, when $\bm{v}(-\infty)=\bm{0}$ it is possible that $s(-\infty)\equiv S$ has a finite value. In that case, the boundary term makes a contribution to Eq. (\ref{2.116}) proportional to
\begin{align}
\lim_{a\rightarrow\infty}\int_0^\infty h(k)\cos ck(t-a)dk,\label{2.118}
\end{align}
where $h(k)=\left[j_1(kb)\right]^2j_1(kS)/k$. However, Eq. (\ref{2.118}) evaluates to zero by the Riemann-Lebesgue lemma \cite{Whittaker}. Hence $B=0$ for all possible $\bm{v}(-\infty)$, so that the force Eq. (\ref{2.116}) is indeed identical to the force Eq. (\ref{2.102}) that was derived by Fourier analysis of the potentials in the Coulomb gauge.
\subsection{Fourier integral and form factor integral \label{2sec3b}}
As discussed above, the self-force Eq. (\ref{2.13}) in terms of a form factor integral is valid for quasi-stationary motion Eq. (\ref{2.31a}) and $\left|\bm{v}\right|\ll c$, and for the special case of harmonic motion. In order to compare Eq. (\ref{2.13}) with the self-force derived in section \ref{2sec2d}, the latter should be specialized accordingly. This may be effected by expanding the integrand of Eq. (\ref{2.101}) in a Taylor series around $t'=t$, and linearizing the result by neglecting all terms nonlinear in $\bm{v}$ and its derivatives. The extremely involved full expansion, in which all nonlinear terms have been kept, is given in Ref. \cite{Roa-Neri1993}. Formally, such use of a Taylor series to describe the integrand on the infinite interval $-\infty<t'<t$ is questionable because the series may have a finite radius of convergence. However, for subrelativistic motion electromagnetic signals are communicated between parts of the charged body on a time scale $\sim b/c$, so that only the small interval $t-b/c\lesssim t'<t$ significantly contributes to the integral in Eq. (\ref{2.101}). This can be seen by noting in Eq. (\ref{2.101}) that the integrand only contributes in the domain $\left|\bm{k}\right|\lesssim b^{-1}$ because the form factor $\left|\rho_k\right|^2\approx 0$ elsewhere, and that the integral over this domain averages out due to the sinusoidal functions unless $ck\tau\lesssim\pi/2$, that is, unless $t-b/c\lesssim t'<t$. Proceeding on this basis by Taylor-expanding, neglecting nonlinear terms, and integrating over angles in $\bm{k}$-space, yields
\begin{align}
\bm{F}=-\frac{8\pi}{3\epsilon_0}\sum_{n=0}^\infty\frac{(-1)^n}{n!}\frac{d^n\bm{v}}{dt^n}\!\!\int_0^\infty\hspace{-2.5mm}\int_0^\infty \hspace{-2mm}k^2\left|\rho_k\right|^2\tau^n\cos ck\tau\,d\tau dk.\label{2.119}
\end{align}
Writing $\tau^{2n}\cos ck\tau=(-1)^nc^{-2n}(d/dk)^{2n}\cos ck\tau$ and\\ $\tau^{2n+1}\cos ck\tau=(-1)^nc^{-2n-1}(d/dk)^{2n+1}\sin ck\tau$, as is suggested in Ref. \cite{Roa-Neri1993}, and integrating by parts with respect to $k$ repeatedly, gives
\begin{align}
\bm{F}=&-\frac{8\pi}{3\epsilon_0}\sum_{n=0}^\infty\frac{(-1)^n}{c^{2n}(2n)!}\frac{d^{2n}\bm{v}}{dt^{2n}}\!\!\int_0^\infty\hspace{-2.5mm}\left(B^{(e)}_n+I^{(e)}_n\right)d\tau\label{2.120}\\
&-\frac{8\pi}{3\epsilon_0}\sum_{n=0}^\infty\frac{(-1)^n}{c^{2n+1}(2n+1)!}\frac{d^{2n+1}\bm{v}}{dt^{2n+1}}\!\!\int_0^\infty\hspace{-2.5mm}\left(B^{(o)}_n+I^{(o)}_n\right)d\tau,\nonumber
\end{align}
in which
\begin{align*}
B^{(e)}_n&=\sum_{m=0}^{2n-1}\left.(-1)^{m}\frac{d^{m}}{dk^{m}}\,k^2\left|\rho_k\right|^2\frac{d^{2n-m-1}}{dk^{2n-m-1}}\cos ck\tau\right|_{k=0}^\infty;\\
B^{(o)}_n&=\sum_{m=0}^{2n}\left.(-1)^{m+1}\frac{d^{m}}{dk^{m}}\,k^2\left|\rho_k\right|^2\frac{d^{2n-m}}{dk^{2n-m}}\sin ck\tau\right|_{k=0}^\infty;\\
I^{(e)}_n&=\int_0^\infty\frac{d^{2n}}{dk^{2n}}\left(k^2\left|\rho_k\right|^2\right)\cos ck\tau\,dk;\\
I^{(o)}_n&=\int_0^\infty\frac{d^{2n+1}}{dk^{2n+1}}\left(k^2\left|\rho_k\right|^2\right)\sin ck\tau\,dk.
\end{align*}
All boundary terms $B^{(e)}_n$ and $B^{(o)}_n$ vanish identically. At $k=\infty$, this is because $\rho_k(\infty)=0$ for any finite charge distribution. At $k=0$, the terms with odd $m$ are zero because $k^2\left|\rho_k\right|^2$ is an even function, and those with even $m$ vanish because they contain $\sin ck\tau$ as a factor. The quantity $\sqrt{2/\pi}I^{(e)}_n\equiv J^{(e)}_n$ can be interpreted as the symmetric cosine transform of the function $j_{\,2n}(k)=(d/dk)^{2n}k^2\left|\rho_k\right|^2$; likewise, $\sqrt{2/\pi}I^{(o)}_n\equiv J^{(o)}_n$ is the symmetric sine transform of $j_{\,2n+1}(k)$. Therefore the double integrals in Eq. (\ref{2.120}) reduce to the single integrals
\begin{align}
\int_0^\infty\hspace{-2mm} I^{(e)}_nd\tau&=\frac{\pi}{2c}\left(\sqrt{\frac{2}{\pi}}\int_0^\infty \hspace{-2mm}J^{(e)}_n\cos kx\,dx\right)_{k=0}\hspace{-2mm}=\frac{\pi}{2c}j_{\,2n}(0);\label{2.125}\\
\int_0^\infty\hspace{-2mm}I^{(o)}_nd\tau&=\frac{1}{c}\sqrt{\frac{2}{\pi}}\left(\int_0^\infty\frac{\sin kx}{k}\,dk\right)\left(\int_0^\infty\hspace{-2mm} J^{(o)}_ndx\right)\label{2.126}\\
&=\int_0^\infty\hspace{-2mm}\left(\sqrt{\frac{2}{\pi}}\int_0^\infty\hspace{-2mm} J^{(o)}_n\sin kx\,dx\right)\!\frac{dk}{ck}=\!\int_0^\infty\hspace{-1mm}\frac{j_{2n+1}(k)}{ck}\,dk.\nonumber
\end{align}
Here, the identity $\int_0^\infty k^{-1}\sin kx\,dk=\pi/2$ and the variable $x=c\tau$ have been used. With the help of Eqs. (\ref{2.125})-(\ref{2.126}), the force Eq. (\ref{2.120}) reduces to
\begin{align}
\bm{F}=\frac{8\pi}{3\epsilon_0c}\sum_{n=0}^\infty\frac{(-1)^{n+1}}{c^{2n}(2n)!}&\left[\frac{\pi}{2}\!\left(\!\frac{d^{2n}}{dk^{2n}}k^2\left|\rho_k\right|^2\right)_{k=0}\frac{d^{2n}\bm{v}}{dt^{2n}}\right.\label{2.127}\\
&\hspace{2mm}\left.+\frac{1}{(2n+1)c}\left(\int_0^\infty\hspace{-2mm}\frac{d^{2n+1}}{dk^{2n+1}}k^2\left|\rho_k\right|^2\frac{dk}{k}\right)\frac{d^{2n+1}\bm{v}}{dt^{2n+1}}\right].\nonumber
\end{align}
This expression now has the manageable form of a series in terms of the derivatives of the current velocity, with coefficients that are readily calculated from the form factor of the charge distribution. In the next section, we will specialize this result to a homogeneously charged sphere, and show that it is equivalent to the series expansion Eq. (\ref{2.32}) obtained by application of Lagrange's reversion theorem. Here, we apply Eq. (\ref{2.127}) to the case of harmonic motion, for which $\bm{v}=-i\omega\bm{\xi}_0\exp(-i\omega t)\equiv-i\omega\widetilde{\bm{\xi}}$. Since $(d/dt)^n\bm{v}=(-i\omega)^n\bm{v}$, Eq. (\ref{2.127}) then becomes the sum of two ordinary power series in the quantity $\omega/c$. The series corresponding to the first line of Eq. (\ref{2.127}) may be interpreted as the even part of the Taylor series of the function $p(\kappa)=\kappa^2\left|\rho_k(\kappa)\right|^2$ around $\kappa=0$, evaluated at $\kappa=\omega/c$. Similarly, the series in the second line may be identified with the odd part of the Taylor series of $p(\kappa)$ around $\kappa=k$, evaluated at $\kappa=k+\omega/c$. Summing these two series therefore results in
\begin{align}
\hspace{-2mm}\bm{F}=\!\frac{4\pi\omega}{3\epsilon_0c}\widetilde{\bm{\xi}}\left(\int_0^\infty\hspace{-3mm}\bigl[p(k+\omega/c)-p(k-\omega/c)\bigr]\frac{dk}{k}\!+\!\frac{i\pi}{2}\bigl[p(\omega/c)+p(-\omega/c)\bigr]\right)\!.\label{2.128}
\end{align}
Noting that $p(k)$ is an even function, the integral in Eq. (\ref{2.128}) may be recognized as the Hilbert transform of $p(k)$ in a less common notation \cite{Zygmund}. Accordingly, by changing variables it may be shown \cite{Zygmund} that
\begin{align}
\bm{F}=\frac{8\pi\omega^2}{3\epsilon_0c^2}\,\widetilde{\bm{\xi}}\!\left(\!\fint_0^\infty\hspace{-1.5mm}\frac{k^2\left|\rho_k(k)\right|^2}{k^2-\omega^2/c^2}\,dk\!+i\pi\underset{k=\omega/c}{\res}\,\frac{k^2\left|\rho_k(k)\right|^2}{k^2-\omega^2/c^2}\right)\!\!,\label{2.129}
\end{align}
where $\fint$ denotes the Cauchy principal value. Here, the second line of Eq. (\ref{2.128}) has been interpreted as a residue. Eq. (\ref{2.129}) is identical to the force Eq. (\ref{2.13}) derived from a Taylor expansion of the retarded integrals for the potentials.
\subsection{Fourier integral and Lagrange expansion \label{2sec3c}}
In the previous section, we derived the series expansion Eq. (\ref{2.127}) that expresses the linearized self-force in terms of the derivatives of the current velocity of the charged body, for a general spherically symmetric charge distribution. We will now specialize this result to a homogeneously charged sphere, and show that this yields the self-force Eq. (\ref{2.32}) that was obtained from series expansion of the retarded potentials. Evaluation of Eq. (\ref{2.127}) using the form factor of a homogeneous sphere Eq. (\ref{2.117}) requires determination of the quantities
\begin{align}
S_n=\left.\frac{d^{2n}\left[j_1(x)\right]^2}{dx^{2n}}\right|_{x=0};\hspace{5mm}T_n=\int_0^\infty\hspace{-1mm}\frac{d^{2n+1}\left[j_1(x)\right]^2}{dx^{2n+1}}\frac{dx}{x}.\label{2.130}
\end{align}
The first of these equals $(2n)!$ times the coefficient of $x^{2n}$ in the Taylor series of $[j_1(x)]^2$ around $x=0$. By squaring the ascending power series of the Bessel function \cite{Abramowitz}, it is thus found that
\begin{align}
S_n=\sum_{m=0}^{n-1}\frac{(2n)!\,\left(-\frac{1}{2}\right)^{n-1}}{m!\,(n-m-1)!\,(2m+3)!!\,(2n-2m+1)!!}.\label{2.131}
\end{align}
Writing factorials in terms of Pochhammer symbols $(p)_q\equiv\Gam(p+q)/\Gam(q)$ with $\Gam$ the Gamma function \cite{Abramowitz}, Eq. (\ref{2.131}) becomes
\begin{align}
S_n=\frac{\pi(2n)!\left(-\frac{1}{4}\right)^{n+1}}{\Gam\left(\frac{5}{2}\right)\Gam(n)\Gam\left(n+\frac{3}{2}\right)}\sum_{m=0}^{n-1}\frac{\left(1-n\right)_m\left(-\frac{1}{2}-n\right)_m}{m!\left(\frac{5}{2}\right)_m}.\label{2.132}
\end{align}
Here, it has been used that $(p)_{-q}=(-1)^q/(1-p)_q$ \cite{Hansen}. The series in Eq. (\ref{2.132}) defines a Gauss hypergeometric function with unit argument \cite{Abramowitz}. Evaluating this hypergeometric function, and converting Gamma functions to factorials, results in
\begin{align}
S_n=-\frac{n(-4)^n}{(n+1)(n+2)(2n+1)}.\label{2.133}
\end{align}
Establishing $T_n$ is more involved. The squared Bessel function $[j_1(x)]^2$ can be expanded in a series of Bessel functions with doubled argument \cite{Watson}. This gives
\begin{align}
\left[j_1(x)\right]^2=\sum_{m=0}^\infty\frac{2m+2}{(2m+1)(2m+3)}\frac{J_{4m+3}(2x)+J_{4m+5}(2x)}{x},\label{2.133}
\end{align}
where $J$ denotes the cylindrical Bessel function of the first kind \cite{Abramowitz}. The factor $x$ in the denominator can be removed with the help of the recurrence relation $2pJ_p(z)/z=J_{p-1}(z)+J_{p+1}(z)$. Subsequently, the integrand of $T_n$ in Eq. (\ref{2.130}) is found by application of the expansion \cite{Luke}
\begin{align}
\frac{d^pJ_q(z)}{dz^p}=\frac{1}{2^p}\sum_{u=0}^p(-1)^u\begin{pmatrix}p\\u\end{pmatrix}J_{q-p+2u}(z).\label{2.134}
\end{align}
This gives
\begin{align}
&\hspace{-2mm}\frac{1}{x}\frac{d^{2n+1}\left[j_1(x)\right]^2}{dx^{2n+1}}=\sum_{m=0}^\infty\sum_{u=0}^{2n+1}\frac{(2m+2)(-1)^u}{(2m+1)(2m+3)}\begin{pmatrix}2n+1\\u\end{pmatrix}\bigg[C_0J_{4m-2n+2u}(2x)\nonumber\\
&\hspace{-2mm}\!+C_2J_{4m-2n+2u+2}(2x)+C_4J_{4m-2n+2u+4}(2x)+C_6J_{4m-2n+2u+6}(2x)\bigg],\label{2.135}
\end{align}
where
\begin{align*}
C_0&=\frac{1}{(4m+3)(4m-2n+2u+1)};\\
C_6&=\frac{1}{(4m+5)(4m-2n+2u+5)};\\
C_2&=C_0+\frac{2(4m+4)}{(4m+3)(4m+5)(4m-2n+2u+3)};\\
C_4&=C_6+\frac{2(4m+4)}{(4m+3)(4m+5)(4m-2n+2u+3)}.
\end{align*}
Substituting this expansion in Eq. (\ref{2.130}), the integral $T_n$ can be evaluated trivially because $\int_0^\infty J_p(z)dz=1$ for arbitrary $p>-1$ \cite{Abramowitz}. Therefore $T_n$ is given by Eq. (\ref{2.135}) if each Bessel function is replaced by $1/2$. The remaining double series can be summed in closed form. The sums over $u$ of the various terms have been tabulated \cite{Hansen}; together they evaluate to
\begin{align}
T_n=&\frac{\pi(-1)^n(2n+1)!}{4\Gam\left(n+\frac{1}{2}\right)\Gam\left(n+\frac{9}{2}\right)}\label{2.136}\\
&\times\sum_{m=0}^\infty\left(4+\frac{1}{m+\frac{1}{2}}-\frac{1}{m+\frac{3}{2}}\right)\!\frac{\left(1\right)_m\left(\frac{1}{4}-\frac{n}{2}\right)_m\left(\frac{3}{4}-\frac{n}{2}\right)_m}{m!\left(\frac{9}{4}+\frac{n}{2}\right)_m\left(\frac{11}{4}+\frac{n}{2}\right)_m}.\nonumber
\end{align}
The series in the second line of Eq. (\ref{2.136}) is derivable from the series
\begin{align}
U_n(z)=\sum_{m=0}^\infty\frac{\left(1\right)_m\left(\frac{1}{4}-\frac{n}{2}\right)_m\left(\frac{3}{4}-\frac{n}{2}\right)_mz^m}{m!\left(\frac{9}{4}+\frac{n}{2}\right)_m\left(\frac{11}{4}+\frac{n}{2}\right)_m},\label{2.137}
\end{align}
which defines the generalized hypergeometric function \cite{Slater}
\begin{align}
U_n(z)=\fourIdx{}{3}{}{2}{F}\!\left[\begin{array}{ccc}1&\frac{1-2n}{4}&\frac{3-2n}{4}\\[3mm]&\frac{11+2n}{4}&\frac{9+2n}{4}\end{array};z\right].\label{2.138}
\end{align}
Comparing Eqs. (\ref{2.136}) and (\ref{2.137}), it is found that
\begin{align}
T_n=\frac{\pi(-1)^n(2n+1)!}{4\Gamma\left(n+\frac{1}{2}\right)\Gamma\left(n+\frac{9}{2}\right)}\!\left(4U_n(1)\!+\hspace{-1mm}\int_0^1\hspace{-1mm}U_n(z)D(z)dz\right)\!,\label{2.139}
\end{align}
with $D(z)=z^{-1/2}-z^{1/2}$. The integral in Eq. (\ref{2.139}) is equal to \cite{Wolfram1}
\begin{align}
\int_0^1\hspace{-1mm}U_n(z)D(z)dz=\frac{4}{3}\,\,\fourIdx{}{4}{}{3}{F}\!\left[\begin{array}{cccc}1&\frac{1}{2}&\frac{1-2n}{4}&\frac{3-2n}{4}\\[3mm]&\frac{5}{2}&\frac{11+2n}{4}&\frac{9+2n}{4}\end{array};1\right].\label{2.140}
\end{align}
Eqs. (\ref{2.138})-(\ref{2.140}) define $T_n$ in terms of two generalized hypergeometric functions of unit argument; for both functions closed form expressions in terms of Gamma functions exist \cite{Wolfram2}. Writing these expressions in terms of factorials yields, after considerable reduction,
\begin{align}
T_n=\frac{\pi(2n+1)(-4)^n}{(n+1)(2n+3)(2n+5)}.\label{2.141}
\end{align}
Finally, having the quantities $S_n$ and $T_n$ at our disposal, the self-force Eq. (\ref{2.127}) can be evaluated. Combining Eqs. (\ref{2.127}), (\ref{2.117}), (\ref{2.133}) and (\ref{2.141}) gives
\begin{align}
\bm{F}=\frac{3q^2}{\pi\epsilon_0cb^2}\sum_{n=0}^\infty&\left(\frac{(2n)(2n+3)(-2b)^{2n}}{(2n+4)!}\frac{1}{c^{2n}}\frac{d^{2n}\bm{v}}{dt^{2n}}\right.\label{2.142}\\
&\hspace{2mm}\left.+\frac{(2n+1)(2n+4)(-2b)^{2n+1}}{(2n+5)!}\frac{1}{c^{2n+1}}\frac{d^{2n+1}\bm{v}}{dt^{2n+1}}\right).\nonumber
\end{align}
Taking the two terms in large braces together by relabeling the summation index gives precisely Eq. (\ref{2.32}). It has thus been shown that the self-force obtained by Fourier analysis of the potentials in the Coulomb gauge is equivalent to the force derived by Lagrange expansion of the potentials in the Lorenz gauge.

\section{Discussion and conclusions \label{2sec4}}
In this paper, we have demonstrated the equivalence of a number of published expressions for the self-force of a rigid charged body. These included the form factor integral Eq. (\ref{2.13}) of Jackson \cite{Jackson} based on a Taylor expansion technique, the series Eqs. (\ref{2.32}) and (\ref{2.33}) of Herglotz \cite{Herglotz} and Schott \cite{Schott} resulting from a Lagrange expansion, the integral over retarded time Eq. (\ref{2.52}) of Sommerfeld \cite{Sommerfeld2} obtained by direct integration of the fundamental equation, and the Fourier integral Eq. (\ref{2.102}) of Bohm and Weinstein \cite{Bohm} derived by Fourier analysis adopting the Coulomb gauge. To this list we may add our result Eq. (\ref{2.127}), which is intermediate between an integral and a series representation.\\

The various expressions differ in their degree of generality. The Fourier type results of Jackson and of Bohm and Weinstein allow calculation of the self-force for arbitrary spherically symmetric charge distributions, whereas the direct integration and Lagrange expansion methods, which stay in the spatial domain throughout, require specialization to a particular charge distribution at an early stage. In addition, several approximations implied in the derivations may be discerned. The neglect of magnetic forces in Eq. (\ref{2.11a}), and of nonlinear terms in the series Eq. (\ref{2.32}), restrict the validity of the corresponding self-forces to nonrelativistic velocities. More subtle is the improper use of the charge distribution at the retarded time, as discussed at the start of section \ref{2sec2b}, which further constrains the results based on Taylor rather than Lagrange expansions to the quasi-stationary motion defined by Eq. (\ref{2.31a}). This restriction does not apply to the more correct self-force expansion Eq. (\ref{2.33}). On the other hand, numerical evaluation of this series by truncation is only possible when the series converges sufficiently rapidly, which is when $(b/c)^{(n+1)}(d/dt)^{(n+1)}\xi\ll(b/c)^n(d/dt)^n\xi$, a condition even stronger than quasi-stationary motion. In contrast, the Fourier technique leading to Eq. (\ref{2.102}) does not involve the explicit evaluation of retarded source densities in the first place, so that the issue is avoided altogether. Therefore, Eq. (\ref{2.102}) may be regarded as the most general self-force expression in this paper, in the sense that this result does not neglect any magnetic or other nonlinear terms, and is not restricted to quasi-stationary motion or any particular charge distribution. At the same time, Eq. (\ref{2.102}) is suitable for numerically stable evaluation as it does not involve truncated series.\\

As emphasized in the introduction of this paper, the electrodynamics of extended charged bodies is important in relation to point charge models of elementary particles, since the two concepts are connected by an appropriate limiting procedure. It is therefore illustrative to see how some well-known point particle results follow from the self-force expressions discussed in this paper. This connection is most straightforward in case of the series expansion Eq. (\ref{2.32}), which is valid for nonrelativistic velocities. In the point particle limit $b\rightarrow0$, the first terms become dominant and evaluate to the well-known nonrelativistic self-force \cite[chap. 21]{Panofsky}
\begin{align}
\bm{F}\rightarrow-\frac{4U_{es}}{3c^2}\frac{d\bm{v}}{dt}+\frac{q^2}{6\pi\epsilon_0c^3}\frac{d^2\bm{v}}{dt^2}+\dots,\label{2.143}
\end{align}
with $U_{es}=3q^2/(20\pi\epsilon_0b)$ the electrostatic energy. The first term on the right of Eq. (\ref{2.143}) has the appearance of an inertial term; correspondingly the quantity $4U_{es}/(3c^2)$ is usually interpreted as an electromagnetic contribution to the mass of the particle. However, the factor $4/3$ violates relativistic mass-energy equivalence, and its removal by more sophisticated arguments has been the subject of much discussion \cite{Janssen,Poincare,RohrlichBoek,Medina,Ori}. The inertial term  is commonly absorbed in an effective mass of the particle, leaving the second term on the right of Eq. (\ref{2.143}) as the radiation reaction force usually featuring in the equation of motion of the classical point particle, in the nonrelativistic limit. The derivation based on Taylor expansion in section \ref{2sec2a} produces the limit Eq. (\ref{2.143}) as well. The first two terms of Eq. (\ref{2.12}) are
\begin{align}
n=0:&\hspace{10mm}-\frac{4}{3c^2}\left(\frac{1}{8\pi\epsilon}\iint\frac{\rho(\bm{r})\rho(\bm{r}')}{\left|\bm{r}-\bm{r}'\right|}d^3\bm{r}d^3\bm{r}'\right)\frac{d\bm{v}}{dt};\label{2.145}\\
n=1:&\hspace{10mm}\frac{1}{6\pi\epsilon_0c^3}{\left(4\pi\!\int_0^\infty\hspace{-2mm}\rho(r)r^2dr\right)\!}^2\frac{d^2\bm{v}}{dt^2}.\label{2.146}
\end{align}
The quantity in large parentheses in the $n=0$ term is just $U_{es}$, so that Eq. (\ref{2.145}) equals the inertial term of Eq. (\ref{2.143}). The quantity in parentheses in the $n=1$ term is the total charge, and Eq. (\ref{2.146}) reduces to the radiation reaction term of Eq. (\ref{2.143}). Moreover, these observations are independent of the particular charge distribution one chooses to model the particle with \cite[chap. 21]{Panofsky}. Less obvious is the point particle limit of the Fourier integral representation of the self-force Eq. (\ref{2.102}), which, as we have shown, is equivalent to the series Eq. (\ref{2.32}), and should therefore reduce to Eq. (\ref{2.143}) as well. However, this is difficult to see from the Fourier integral directly. In contrast, from our intermediate result Eq. (\ref{2.127}) the limit Eq. (\ref{2.143}) is readily obtained. Namely, the term $n=0$ in the first line of Eq. (\ref{2.127}) vanishes for any finite $\rho_k$. The term $n=0$ in the second line gives, after integration by parts, the force
\begin{align}
-\frac{2}{3\epsilon_0c^2}\frac{d\bm{v}}{dt}\int\frac{\left|\rho_k\right|^2}{k^2}\,d^3\bm{k}.\label{2.144}
\end{align}
By writing out $\rho_k$ as the Fourier transform of the charge distribution $\rho(\bm{r})$, and simplifying the result by using the identity $\int\exp\left[-i\bm{k}\cdot(\bm{r}-\bm{r}')\right]k^{-2}d^3\bm{k}=2\pi^2/\left|\bm{r}-\bm{r}'\right|$ \cite{Morse}, Eq. (\ref{2.144}) is reduced to Eq. (\ref{2.145}) and hence to the inertial term of Eq. (\ref{2.143}). The radiation reaction force, in turn, is generated by the term $n=1$ in the first line of Eq. (\ref{2.127}). Exploiting in the latter spherical symmetry by writing $\rho_k=\sqrt{2/\pi}\int_0^\infty\rho(r)j_0(kr)r^2dr$, performing the differentiations with respect to $k$ and evaluating the result for $k=0$, yields the force Eq. (\ref{2.146}). None of the other terms of Eq. (\ref{2.127}) contribute in the point particle limit. This is because $\rho_k$ will tend to a constant as the support of $\rho(\bm{r})$ shrinks to a point, and therefore the differentiations with respect to $k$ will make these terms vanish. This confirms that also the Fourier integral Eq. (\ref{2.127}) correctly reduces to the well-known self-force Eq. (\ref{2.143}).\\

An important aspect of the relativistic dynamics of an extended body, left out of the discussion so far, is the way is which the body maintains or changes its shape while being accelerated. On the one hand relativity theory requires a velocity dependent Lorentz contraction; on the other hand instantaneous contractions are impossible due to the finite velocity at which information about velocity changes can propagate through the extended body. Accelerated motion, therefore, necessitates some notion of relativistic rigidity, the precise formulation of which is still being studied \cite{Lyle,Epp}. Often the approximation of Born rigidity \cite{Born} is applied, valid for adiabatic velocity changes, where it is assumed that the accelerated body always maintains its shape in its continuously changing instantaneous rest frame \cite{Pierce}. The corresponding Born rigid extended electron model was advocated by Lorentz \cite{Lorentz}. A significant strength of this model is that it provides a natural way to cure the 4/3 problem described above, by assuming a negative pressure inside the electron that balances the Coulomb repulsion of the distributed charge \cite{Poincare}. The work done by this pressure during Lorentz contractions removes the factor 4/3 precisely \cite{Yaghjian,Medina}. (Another possibility is to adopt a manifestly covariant definition of four-momentum of systems involving electromagnetic fields \cite{RohrlichBoek,Ori,Janssen}). However, none of the self-force derivations discussed in this paper corresponds to the Lorentz model. Rather, the Abraham model \cite{Abraham} is implied, in which the charged body maintains its shape in the frame of an observer at rest. Differences between the self-forces according to each model do not yet appear in the linear terms, however, so that the limit Eq. (\ref{2.143}) is model independent. In the Lagrange series Eq. (\ref{2.33}), on the other hand, the sum over $m$ of the nonlinear terms $n=0$ indeed agrees \cite{Schott} with the model of Abraham and not with that of Lorentz. Similarly, the time integral Eq. (\ref{2.50}), and therefore also the equivalent Fourier integral Eq. (\ref{2.102}), coincide with the Abraham model \cite{Sommerfeld1}. Since the contracting Lorentz model seems more in line with special relativity, it would be valuable for classical charged particle theories to derive a Born rigid version of the Fourier representation Eq. (\ref{2.102}) of the self-force. A covariant formulation based on formal cut-off procedures has been developed in this direction \cite{Prigogine}. Yet, it should be realized that it has not been established that a consistent extended body model of elementary particles, if any, should necessarily respect rigidity. Therefore alternative postulates, such as the Abraham model or other non-rigid models \cite{Aguirregabiria}, may continue to prove their value. What is more, the current state of technology is starting to enable experimental conditions in which the electromagnetic self-force of macroscopic charged systems, such as high-density electron bunches \cite{Smorenburg1} and ultracold plasma bunches \cite{Smorenburg2}, becomes significant. It would be interesting to see to what extent the self-force formulations in this paper can model these evidently non-rigid systems.

\begin{acknowledgements}
This work is part of the research program of the Foundation for Fundamental Research on Matter (FOM), which is part of the Netherlands Organization for Scientific Research (NWO). 
\end{acknowledgements}

\hyphenation{E-lek-tro-nen-selbst-be-schleu-ni-gung}

\end{document}